\documentclass[aps,preprint]{revtex4}%
\usepackage{amsfonts}
\usepackage{amsmath}
\usepackage{amssymb}
\usepackage{graphicx}%
\begin{document}
\title[Center of Energy]{Illustrations of the Relativistic Conservation Law for the Center of Energy}
\author{Timothy H. Boyer}
\affiliation{Department of Physics, City College of the City University of New York, New
York, NY 10031}

\begin{abstract}
The relativistic conservation law involving the center of energy is reviewed
and illustrated using simple examples from classical electromagnetic theory.
\ It is emphasized that this conservation law is parallel to the conservation
laws for energy, linear momentum, and energy, in arising from the generators
of the Poincare group for electromagnetic theory; yet this relativistic law
reflecting the continuous flow of energy goes virtually unmentioned in the
text books. \ The illustrations here present situations both where external
forces are present and are absent. \ The cases of a parallel plate capacitor,
a flattened slip-joint solenoid, and two interacting charges are included.

\end{abstract}
\maketitle

\section{INTRODUCTION}

Classical electrodynamics, like any other relativistic Lagrangian field
theory, is invariant under the Poincare group involving the operations of
spacetime translation, spatial rotation, and proper Lorentz transformation.
\ The associated infinitesimal generators,\cite{ColeVan} designated by
$\mathbf{P}$, $U$, $\mathbf{L}$, and $U\overrightarrow{\mathcal{X}}$, are
associated with conserved quantities. \ The generator $\mathbf{P}$ of space
translations is associated with conservation of linear momentum. \ The
generator $U$ of time translations is associated with conservation of energy.
\ The generator $\mathbf{L}$ of spatial rotations is associated with
conservation of angular momentum. \ The generator $U\overrightarrow
{\mathcal{X}}$ of proper Lorentz transformations is associated with the
uniform motion of the system center of energy.\cite{CofE} \ Although the
conservation laws of linear momentum, angular momentum, and energy are
illustrated by fine elementary examples in electromagnetism text
books,\cite{Griffiths} this does not seem to be the case for the uniform
motion of the center of energy.\cite{Examples} \ The invariant motion of the
center of energy is well\cite{Einstein} but not widely known, and is rarely
illustrated with examples in the electromagnetism literature.\ The law
expresses the continuous flow of energy in relativistic systems. \ In this
article we review the relativistic law for the invariant motion of the center
of energy and then present three simple electromagnetic examples: a
parallel-plate capacitor, a flattened, slip-joint solenoid, and two
interacting point charges. \ The examples remind us that when calculating the
center of energy of an electromagnetic system, relativistic particle equations
of motion must be used and all the energy must be considered, including the
particle rest energy and kinetic energy, and the distributed energy stored in
the electromagnetic field.

\section{RELATIVISTIC CONSERVATION LAWS}

\subsection{The Generators of the Poincare Group for Electromagnetism}

For charged point masses $m_{i}$ interacting through electromagnetic fields
$\mathbf{E}$ and $\mathbf{B}$, the generators of the Poincare
group\cite{ColeVan} take the forms%
\begin{equation}
\mathbf{P}=%
{\displaystyle\sum_{i}}
m_{i}\gamma_{i}\mathbf{v}_{i}+%
{\displaystyle\int}
d^{3}r\,\frac{1}{4\pi c}\mathbf{E}\times\mathbf{B}\text{
\ \ \ \ \ \ \ \ \ \ \ \ \ \ \ \ \ \ \ \ (linear momentum)}%
\end{equation}%
\begin{equation}
U=%
{\displaystyle\sum_{i}}
m_{i}\gamma_{i}c^{2}+%
{\displaystyle\int}
d^{3}r\,\frac{1}{8\pi}(E^{2}+B^{2})\text{
\ \ \ \ \ \ \ \ \ \ \ \ \ \ \ \ \ \ (energy)}%
\end{equation}%
\begin{equation}
\mathbf{L}=%
{\displaystyle\sum_{i}}
\mathbf{r}_{i}\times m_{i}\gamma_{i}\mathbf{v}_{i}+%
{\displaystyle\int}
d^{3}r\,\mathbf{r}\times\left(  \frac{1}{4\pi c}\mathbf{E}\times
\mathbf{B}\right)  \text{ \ \ \ \ \ \ \ \ \ \ \ \ \ \ \ \ \ \ \ \ \ (angular
momentum)}%
\end{equation}
and%
\begin{equation}
U\overrightarrow{\mathcal{X}}=%
{\displaystyle\sum_{i}}
\mathbf{r}_{i}m_{i}\gamma_{i}c^{2}+%
{\displaystyle\int}
d^{3}r\,\mathbf{r}\,\frac{1}{8\pi}(E^{2}+B^{2})\text{ \ \ \ \ \ \ \ (energy
times center of energy)}%
\end{equation}
where $\mathbf{v}_{i}=d\mathbf{r}_{i}/dt$ is the time-derivative of the
particle displacement $\mathbf{r}_{i}$, $\gamma_{i}=(1-v_{i}^{2}/c^{2}%
)^{-1/2}$, $\mathbf{E}$ and $\mathbf{B}$ represent the electric and magnetic
fields, and $\overrightarrow{\mathcal{X}}$ is the center of energy of the
system. \ These correspond to the generators respectively of space
translation, time translation, spatial rotation, and proper Lorentz
transformation. \ In the absence of external forces, the first three
quantities are time-independent and the fourth has a constant time derivative.
\ The electromagnetic expressions in the first three equations appear in the
electromagnetism textbooks, whereas the last is usually absent. \ On account
of this omission, we will sketch the derivation of the center of energy expression.

\subsection{Derivation of the Center-of-Energy Law}

The center of energy $\overrightarrow{\mathcal{X}}$ in Eq. (4) is analogous to
the familiar center of (rest) mass $\overrightarrow{\mathcal{X}}_{restmass}$
of nonrelativistic mechanics%
\begin{equation}
M\overrightarrow{\mathcal{X}}_{restmass}=%
{\displaystyle\sum\limits_{i}}
m_{i}\mathbf{r}_{i}\text{ ,\ \ \ \ \ }M=%
{\displaystyle\sum\limits_{i}}
m_{i}%
\end{equation}
except that all energy contributes. \ The total energy $U$ in Eq. (2) for a
system of charged particles and electromagnetic fields is the sum of the
relativistic mechanical energy of each particle $m_{i}\gamma_{i}c^{2}$ and the
electromagnetic field energy found by integrating the energy density
$u=[1/(8\pi)](E^{2}+B^{2})$ over all space. \ The center-of-energy expression
(4) involves weighting the displacement $\mathbf{r}$ by the amount of the
energy located at $\mathbf{r}$. \ Thus a point mass of energy $m_{i}\gamma
_{i}c^{2}$ contributes $\mathbf{r}_{i}(m_{i}\gamma_{i}c^{2})$ while the
electromagnetic energy $u\,d^{3}r$ in a differential volume $d^{3}r$
contributes $\mathbf{r}(u\,d^{3}r)=\mathbf{r}[1/(8\pi)](E^{2}+B^{2})d^{3}r$.
\ Summing over the particles and integrating over all the electromagnetic
fields in space, we obtain the expression (4) for the energy times the center
of energy $U\overrightarrow{\mathcal{X}}$.

The derivation of the law for the invariant motion of the center of energy in
electromagnetic theory can be given in a fashion parallel to that given for
Poynting's theorem.\cite{Poynting} \ We consider the integral over all space
of $%
{\displaystyle\int}
d^{3}r\,\mathbf{r\,}(\mathbf{J\cdot E)}$ which represents the volume-integral
over the displacement $\mathbf{r}$ weighted by $\mathbf{J\cdot E}$, the local
transfer of power from electromagnetic form over to some other form due to the
forces produced by electric fields on moving charges. \ Just as for Poynting's
theorem, we use Maxwell's equations to rewrite this integral in terms of the
electromagnetic fields alone,
\begin{align}%
{\displaystyle\int}
d^{3}r\,\mathbf{r\,}(\mathbf{J\cdot E)}  &  \mathbf{=}%
{\displaystyle\int}
d^{3}r\,\mathbf{r}\left[  \frac{c}{4\pi}(\nabla\times\mathbf{B-}\frac{1}%
{c}\frac{\partial\mathbf{E}}{\partial t})\cdot\mathbf{E}\right] \nonumber\\
&  =%
{\displaystyle\int}
d^{3}r\,\mathbf{r}\left[  \frac{-c}{4\pi}\nabla\cdot(\mathbf{E}\times
\mathbf{B})+\frac{c}{4\pi}\mathbf{B}\cdot(\nabla\times\mathbf{E})-\frac
{1}{8\pi}\frac{\partial}{\partial t}E^{2}\right] \nonumber\\
&  =%
{\displaystyle\int}
d^{3}r\,\mathbf{r}\left[  -\nabla\cdot\left(  \frac{c}{4\pi}\mathbf{E}%
\times\mathbf{B}\right)  -\frac{\partial}{\partial t}\left(  \frac{1}{8\pi
}(E^{2}+B^{2})\right)  \right] \nonumber\\
&  =-%
{\displaystyle\int}
\mathbf{r}\,\left(  \frac{c}{4\pi}\mathbf{E}\times\mathbf{B}\right)  \cdot
d\mathbf{A+}%
{\displaystyle\int}
d^{3}r\left(  \frac{c}{4\pi}\mathbf{E}\times\mathbf{B}\right)  -\frac{d}{dt}%
{\displaystyle\int}
d^{3}r\,\mathbf{r}\,\left(  \frac{1}{8\pi}(E^{2}+B^{2})\right) \nonumber\\
&  =%
{\displaystyle\int}
d^{3}r\left(  \frac{c}{4\pi}\mathbf{E}\times\mathbf{B}\right)  -\frac{d}{dt}%
{\displaystyle\int}
d^{3}r\,\mathbf{r}\,\left(  \frac{1}{8\pi}(E^{2}+B^{2})\right)
\end{align}
where we have used the divergence theorem and have dropped the surface term
assuming that the sources of electromagnetic fields are localized.

For a system of charged particles interacting through the electromagnetic
fields, we differentiate Eq. (4) to obtain
\begin{align}
\frac{d(U\overrightarrow{\mathcal{X}}\mathbf{)}}{dt}  &  =\frac{d}{dt}\left(
{\displaystyle\sum}
\mathbf{r}_{i}m_{i}\gamma_{i}c^{2}+%
{\displaystyle\int}
d^{3}r\,\mathbf{r}\frac{1}{8\pi}\,(E^{2}+B^{2})\right) \nonumber\\
&  =c^{2}%
{\displaystyle\sum}
m_{i}\gamma_{i}\mathbf{v}_{i}+%
{\displaystyle\sum}
\mathbf{r}_{i}\frac{d}{dt}(m_{i}\gamma_{i}c^{2})+\frac{d}{dt}%
{\displaystyle\int}
d^{3}r\,\mathbf{r}\,\frac{1}{8\pi}(E^{2}+B^{2})\nonumber\\
&  =c^{2}%
{\displaystyle\sum}
m_{i}\gamma_{i}\mathbf{v}_{i}+%
{\displaystyle\int}
d^{3}r\,\mathbf{r(J\cdot E)+}\frac{d}{dt}%
{\displaystyle\int}
d^{3}r\,\mathbf{r}\,\frac{1}{8\pi}(E^{2}+B^{2})\nonumber\\
&  =c^{2}\left[
{\displaystyle\sum}
m_{i}\gamma_{i}\mathbf{v}_{i}+%
{\displaystyle\int}
d^{3}r\left(  \frac{1}{4\pi c}\mathbf{E}\times\mathbf{B}\right)  \right]
=c^{2}\mathbf{P}%
\end{align}
where we have used the result of Eq. (6) and the energy transfer equation for
point charges
\begin{equation}%
{\displaystyle\sum}
\mathbf{r}_{i}\frac{d}{dt}(m_{i}\gamma_{i}c^{2})=%
{\displaystyle\int}
d^{3}r\,\mathbf{r(J.E)}%
\end{equation}
where the point-charge current density is\ $\mathbf{J(r},t)\mathbf{=}%
{\displaystyle\sum}
q_{i}\mathbf{v}_{i}\delta^{3}(\mathbf{r-r}_{i}(t))$ and $d(m_{i}\gamma
_{i}c^{2})/dt=q_{i}\mathbf{v}_{i}\cdot\mathbf{E(r}_{i},t)$. \ Thus in Eq. (7)
we see that the time rate of change of the quantity \{energy times the center
of energy\} is equal to $c^{2}$ times the linear momentum of the system.
\ Since the linear momentum and the energy of the system are constant in time,
this means that the velocity of the center of energy is constant in time,
$d\overrightarrow{\mathcal{X}}/dt=const.$

\subsection{Conservation Laws in the Presence of External Forces on Particles}

In many cases it is convenient to consider not isolated electromagnetic
systems but rather electromagnetic systems in interaction with external forces
$\mathbf{F}_{ext\text{ }i}$ acting on the particles of the system. \ In this
case the conservation ideas are changed for all the conservation laws. \ The
the sum of the external forces gives the time-rate-of-change of the system
linear momentum%
\begin{equation}%
{\displaystyle\sum\limits_{i}}
\mathbf{F}_{ext\text{ }i}=\frac{d\mathbf{P}}{dt}%
\end{equation}
The power delivered by the external forces gives the time-rate-of-change of
the system energy%
\begin{equation}%
{\displaystyle\sum\limits_{i}}
\mathbf{F}_{ext\text{ }i}\cdot\mathbf{v}_{i}=\frac{dU}{dt}%
\end{equation}
The sum of the external torques gives the time-rate-of-change of the system
angular momentum $\mathbf{L}$ (about the origin)
\begin{equation}%
{\displaystyle\sum\limits_{i}}
\mathbf{r}_{i}\times\mathbf{F}_{ext\text{ }i}=\frac{d\mathbf{L}}{dt}%
\end{equation}
The law for the change in the energy times the center of energy seems
unfamiliar.\cite{unaware}\ \ We can obtain the rule by using the modified
equation of energy transfer $d(m_{i}\gamma_{i}c^{2})/dt=[q_{i}\mathbf{E(r}%
_{i},t)+\mathbf{F}_{ext\,i}]\cdot\mathbf{v}_{i}$ for the $i$-th particle
(multiplied by $\mathbf{r}_{i}$)%
\begin{equation}
\mathbf{r}_{i}\frac{d}{dt}(m_{i}\gamma_{i}c^{2})=\mathbf{r}_{i}(\mathbf{F}%
_{ext\text{ }i}\cdot\mathbf{v}_{i})+\mathbf{r}_{i}(q_{i}\mathbf{E}%
\cdot\mathbf{v}_{i})
\end{equation}
and summing over all the particles%
\begin{equation}%
{\displaystyle\sum\limits_{i}}
\mathbf{r}_{i}\frac{d}{dt}(m_{i}\gamma_{i}c^{2})=%
{\displaystyle\sum\limits_{i}}
\mathbf{r}_{i}(\mathbf{F}_{ext\text{ }i}\cdot\mathbf{v}_{i})+%
{\displaystyle\int}
d^{3}r\,\mathbf{r}(\mathbf{J}\cdot\mathbf{E})
\end{equation}
Now using Eq. (13) for $%
{\displaystyle\int}
d^{3}r\,\mathbf{r}(\mathbf{J}\cdot\mathbf{E})$ in Eq. (6) and noting the first
two lines of Eq. (7), we obtain the rule for the center of energy,%
\begin{equation}%
{\displaystyle\sum\limits_{i}}
(\mathbf{F}_{ext\text{ }i}\cdot\mathbf{v}_{i})\,\mathbf{r}_{i}=\frac
{d(U\overrightarrow{\mathcal{X}})}{dt}-c^{2}\mathbf{P}%
\end{equation}
Thus the power weighted by the position where the power is delivered equals
the time-rate-of-change of the system energy times the center of energy minus
$c^{2}$ times the system linear momentum. All of the laws (9)-(11), (14) can
be integrated with respect to time so as to give integral forms. \ The
integral form for the relativistic center-of-energy law in (14) is%

\begin{equation}%
{\displaystyle\sum\limits_{i}}
{\displaystyle\int_{1}^{2}}
(d\mathbf{r}_{i}\cdot\mathbf{F}_{ext\text{ }i})\,\mathbf{r}_{i}=U_{2}%
\overrightarrow{\mathcal{X}}_{2}-U_{1}\overrightarrow{\mathcal{X}}_{1}-c^{2}%
{\displaystyle\int_{1}^{2}}
dt\mathbf{P}%
\end{equation}
\ In special relativity, the flow of energy has a continuous meaning. \ Thus
the introduction of energy by external forces located at points in space
changes the center of energy of the system. \ 

The continuous flow of energy in space for relativistic systems is in contrast
with the situation in nonrelativistic mechanics where energy can be suddenly
transported from one point in space to another. \ Thus in nonrelativistic
mechanics a long, massless, rigid pole can be used to transport energy
instantaneously from one end of the pole to the other. \ Such poles do not
exist in relativistic physics. \ Rather, in relativistic physics a system has
a well-defined center of energy which moves through space continuously at a
speed (in the absence of external forces) not exceeding the speed of light in
vacuum $c$.

It is interesting to note the nonrelativistic limit for the center-of-energy
relations in Eqs. (4), (7) and (14). \ If we divide by a factor of $c^{2}$ and
allow $c\rightarrow\infty$, then all that remains of the energy given in Eq.
(2) is the rest-mass contribution $U/c^{2}$ $\rightarrow%
{\displaystyle\sum\limits_{i}}
m_{i}$ with no contribution from the (finite) kinetic energy or
electromagnetic energy. \ Thus in the $c\rightarrow\infty$ limit, Eq. (4)
becomes the expression for the center of rest mass given in Eq. (5). \ Also,
equation (7) becomes the statement that the total rest mass times the center
of rest mass equals the momentum%
\begin{equation}
\frac{d}{dt}\left[  \left(
{\displaystyle\sum\limits_{i}}
m_{i}\right)  \overrightarrow{\mathcal{X}}\right]  =%
{\displaystyle\sum\limits_{i}}
m_{i}\mathbf{v}_{i}=\mathbf{P}\text{ \ \ \ \ \ \ (}c\rightarrow\infty\text{
limit)}%
\end{equation}
These are familiar results in Galilean-invariant (nonrelativistic) mechanics.
\ On dividing Eq. (14) by $c^{2}$ and allowing $c\rightarrow\infty$, the
left-hand side involving external forces vanishes entirely and the right-hand
side involves simply the same statement in Eq. (16) obtained from the
$c\rightarrow\infty$ limit of Eq. (7). \ Within nonrelativistic physics, there
is a continuous flow of rest mass but not of energy. \ Thus in nonrelativistic
physics there is no separate law regarding the \textit{location} where energy
is introduced into the system.

\section{ILLUSTRATIONS OF THE CENTER-OF-ENERGY CONSERVATION LAW}

\subsection{Quasi-static Changes for Stationary Systems}

\subsubsection{A Single Point Mass}

\ \ \ \ As the simplest possible example of the relativistic conservation laws
for stationary systems, we consider a single point mass $m$ at rest at
displacement $\mathbf{r}$ in some inertial frame. \ The conserved quantities
associated with Poincare invariance then involve energy $U=mc^{2}$, linear
momentum $P=0$, angular momentum about the origin $L=0$, and energy times
center of energy $(mc^{2})\overrightarrow{\mathcal{X}}=mc^{2}\mathbf{r}$. \ We
now use an external force $\mathbf{F}_{ext}$ to move the mass from
$\mathbf{r}$ to $\mathbf{r}^{\prime}$ quasi-statically. \ Since the external
force can be chosen arbitrarily small, there is no linear impulse delivered,
no net work done, no angular impulse, and no moment-of-work done. \ The only
conservation law with some non-vanishing terms is the fourth involving the
center of energy. \ Here the system linear momentum has a nonvanishing
time-intergral so that the integral form of the law in Eq. (15) gives
\begin{equation}
0=(mc^{2}\mathbf{r}^{\prime})-(mc^{2}\mathbf{r})-c^{2}%
{\displaystyle\int\limits_{1}^{2}}
\mathbf{P}\,dt
\end{equation}
which is consistent with the momentum of a particle%
\begin{equation}
\mathbf{P}=\frac{m}{\sqrt{1-[(d\mathbf{r}/dt)/c]^{2}}}\frac{d\mathbf{r}}%
{dt}\cong m\frac{d\mathbf{r}}{dt}\text{ \ \ \ \ \ \ \ (quasi-static)}%
\end{equation}
We notice that even though the linear momentum $\mathbf{P}$ can be made as
small as desired by taking the external forces sufficiently small, the time
integral of the linear momentum gives a finite non-zero value independent of
the magnitude of the small external force in the limit $d\mathbf{r}%
/dt\rightarrow0$. \ The change in the position of the system center of energy
was associated with a flow of momentum as required by special relativity.

In this simplest case where all of the energy is rest-mass energy, we could
actually have divided Eq. (17) through by $c^{2}$ and have obtained a result
valid in nonrelativistic physics where the linear momentum is given by exactly
$\mathbf{p}=m\mathbf{v}$. \ In nonrelativistic physics, the change in
\textit{rest} mass position is continuous and is associated with the flow of
linear momentum.

\subsubsection{Parallel Plate Capacitor}

A parallel-plate capacitor provides a simple illustration of the conservation
law for the center of energy when electrostatic energy is involved. \ The
electrostatic energy contributes to the center of energy of the system in
relativistic physics, whereas it does not contribute to the center of rest
mass which appears in nonrelativistic physics. \ \ We consider a capacitor
consisting of two parallel conducting plates, each of dimension $L\times L$,
the left-hand plate of mass $m$ in the plane with $x$-coordinate $x,$ and the
right-hand plate of mass $M$ in the plane with $x$-coordinate $X$. \ In this
section discussing quasi-static displacement, we will take the masses $m$ and
$M$ as negligible. \ The plates are centered so that the x-axis passes through
the center of each plate. Plate $m$ is charged with total charge $+Q$ and
plate $M$ with charge $-Q$. \ It is assumed that the plates form a
parallel-plate capacitor of small separation $0<X-x<<L$ with an electric field
given by the electrostatic expression
\begin{equation}
\mathbf{E=}\widehat{i}4\pi Q/L^{2}%
\end{equation}
between the plates. \ There is no magnetic field present and it is assumed
that we may neglect the fringing fields outside the plates.

In order to maintain the capacitor plates at rest, there must be external
forces\cite{Onehalf} $F_{extm}=-Q(E+0)/2=-2\pi Q^{2}/L^{2}=-F_{extM}$ on the
left-hand plate of negligible rest mass $m$ at $x$ and on the right-hand plate
of negligible rest mass $M$ at $X$ respectively. \ The illustration of energy
conservation for this situation is easily carried out.\cite{solenoid} \ Thus
if the two plates are displaced quasi-statically from $x$ to $x^{\prime}$ and
from $X$ to $X^{\prime}$ respectively, the work done by the external forces of
constraint is found to equal the change in electrostatic energy%

\begin{align}
F_{extm}(x^{\prime}-x)+F_{extM}(X^{\prime}-X) &  =\frac{2\pi Q^{2}}{L^{2}%
}[-(x^{\prime}-x)+(X^{\prime}-X)]\nonumber\\
&  =\frac{1}{8\pi}\left(  \frac{4\pi Q}{L^{2}}\right)  ^{2}L^{2}(X^{\prime
}-x^{\prime})-\frac{1}{8\pi}\left(  \frac{4\pi Q}{L^{2}}\right)  ^{2}%
L^{2}(X-x)]
\end{align}
However, in contrast to the work-energy law, the relativistic center-of-energy
law in Eq. (15) usually goes unmentioned. \ On quasi-static displacement of
the plates, there is no magnetic field generated in the region between the
plates and therefore no electromagnetic field momentum between the plates.
\ If the plates are displaced quasi-statically from $x$ to $x^{\prime}$ and
from $X$ to $X^{\prime}$, then the left-hand side of Eq. (15)\ gives
\begin{align}%
{\displaystyle\sum\limits_{i}}
{\displaystyle\int}
dx_{i}F_{extxi}x_{i} &  =%
{\displaystyle\int\limits_{x}^{x^{\prime}}}
dx^{\prime\prime}\left(  \frac{-2\pi Q^{2}}{L^{2}}\right)  x^{\prime\prime}+%
{\displaystyle\int\limits_{X}^{X^{\prime}}}
dX^{\prime\prime}\left(  \frac{2\pi Q^{2}}{L^{2}}\right)  X^{\prime\prime
}\nonumber\\
&  =\frac{-\pi Q^{2}(x^{\prime2}-x^{2})}{L^{2}}+\frac{\pi Q^{2}(X^{\prime
2}-X^{2})}{L^{2}}%
\end{align}
while the right-hand side of Eq. (15) gives%
\begin{equation}
U_{2}\mathcal{X}_{2}-U_{1}\mathcal{X}_{1}-c^{2}%
{\displaystyle\int\limits_{1}^{2}}
dtP_{x}=\frac{\pi Q^{2}(X^{\prime2}-x^{\prime2})}{L^{2}}-\frac{\pi Q^{2}%
(X^{2}-x^{2})}{L^{2}}-0
\end{equation}
After rearrangement, equations (21) and (22) are seen to involve the same
quantities on the right-hand sides. \ Thus indeed moving the capacitor plates
illustrates the relativistic center-of-energy law (15) with external forces.
\ Thus the energy introduced by the external forces at the plates provides not
only the change in electrostatic energy but also the continuous motion of the
center of electrostatic energy.

\subsubsection{Flattened, Slip-Joint Solenoid}

It was pointed out recently\cite{solenoid} that energy calculations for a
solenoid can be made analogous to those for a parallel-plate capacitor by
flattening the solenoid and fitting it with slip joints which allow relative
motion of the front and back current sheets while maintaining the continuity
of the circulating surface currents. \ Here we will use this solenoidal
configuration to carry out calculations for a solenoid which are analogous to
those given above for a capacitor.

The flattened solenoid consists of two large perfectly-conducting plates of
size $L\times l$ with negligible masses $m$ and $M$ located in the planes $x$
and $X$ respectively and connected through short perfectly-conducting sides
parallel to the $yz$-plane which are fitted with slip joints. \ The slip
joints maintain the continuity of the electrical circuit while allowing the
plates to move along the $x$-axis, which passes through the centers of the
plates. \ The surface current $\mathbf{K}$ is always perpendicular to the
$\widehat{k}$-direction and flows around the solenoid, in the $+\widehat{j}$
direction in the $M$ plate and in the $-\widehat{j}$ direction in the $m$
plate. \ The surface current $\mathbf{K}$ causes a magnetic field
\begin{equation}
\mathbf{B}=\widehat{k}4\pi K/c
\end{equation}
parallel to the $z$-axis within the flattened solenoid. \ The magnetic flux
$\Phi$ through the solenoid is given by the magnitude of $\mathbf{B}$ times
the cross-sectional area
\begin{equation}
\Phi=BL(X-x)=4\pi KL(X-x)/c
\end{equation}
\ We assume that the separation between the plates is very small $0<X-x<<L,l$
compared to the other dimensions so that we can neglect the fringing fields.

\ Here we are interested in the case where external mechanical forces on the
left and right current sheets of the flattened solenoid allow these to change
location quasi-statically from $x$ to $x^{\prime}$ and from $X$ to $X^{\prime
}$ respectively. \ We assume that there is no ohmic resistance in the sheets
nor any batteries present, so that the currents of the solenoid flow in such a
fashion as to maintain the total magnetic flux $\Phi$ through the solenoid as
constant in time. \ The external forces needed to balance the magnetic forces
on the current sheets at $x$ and $X$ are given by
\begin{equation}
\mathbf{F}_{extx}=\widehat{i}\frac{KLl}{c}\frac{(\mathbf{B}+0)}{2}=\widehat
{i}\frac{1}{8\pi}B^{2}Ll=\widehat{i}\frac{1}{8\pi}\frac{\Phi^{2}l}{L(X-x)^{2}%
}=-\mathbf{F}_{extX}%
\end{equation}
while the energy in the magnetic field is given by
\begin{equation}
U=\frac{1}{8\pi}B^{2}Ll(X-x)=\frac{1}{8\pi}\frac{\Phi^{2}l}{L(X-x)}%
\end{equation}
Just as in the electrostatic case, it is easy to verify the connection between
the external forces and the energy changes for the solenoid. \ In this case,
we use the differential form in Eq. (10), finding%
\begin{equation}
\mathbf{F}_{extx}\cdot\widehat{i}\frac{dx}{dt}+\mathbf{F}_{extX}\cdot
\widehat{i}\frac{dX}{dt}=\frac{1}{8\pi}\frac{\Phi^{2}l}{L(X-x)^{2}}\left(
\frac{dx}{dt}-\frac{dX}{dt}\right)  =\frac{d}{dt}\left(  \frac{1}{8\pi}%
\frac{\Phi^{2}l}{L(X-x)}\right)  =\frac{dU_{em}}{dt}%
\end{equation}
which confirms the energy conservation law.

It is also possible to verify the law in Eq. (14) for the relativistic center
of energy. \ Now since the cross-sectional area of the solenoid is changing,
it follows that the magnetic field must be changing and this means that
electric fields must be induced. \ Induced electric fields together with the
solenoid magnetic field will lead to electromagnetic field linear momentum and
hence to a contribution in Eq. (14) from $c^{2}\mathbf{P}$. \ In order to find
the electric field induced when the current sheets are moved apart, we
consider a single current sheet seen in a new Lorentz frame. \ If we consider
a current sheet normal to the $x$-axis with a current $\mathbf{K}$ flowing in
the $\widehat{j}$ direction, then there is a magnetic field $\mathbf{B}%
=\pm\widehat{k}(2\pi/c)\mathbf{K}$, the factor of $2\pi$ rather than $4\pi$
since only a single current sheet is involved. \ Under Lorentz transformation
to a new inertial frame moving with velocity $v=c\beta$ along the x-axis, one
finds a uniform electric field $\mathbf{E}=\pm\widehat{j}\gamma\beta B\cong
\pm\widehat{j}\gamma(v/c)(2\pi K/c)$. \ Applying this to both plates of the
capacitor, we find that in the region between the moving current sheets, there
is a net electric field%
\begin{equation}
\mathbf{E}=\widehat{j}\gamma_{m}\frac{2\pi K}{c^{2}}\frac{dx}{dt}+\widehat
{j}\gamma_{M}\frac{2\pi K}{c^{2}}\frac{dX}{dt}%
\end{equation}
Thus inside the flattened solenoid, there is an electromagnetic linear
momentum%
\begin{align}
\mathbf{P}  &  =\frac{1}{4\pi c}\mathbf{E}\times\mathbf{B}Ll(X-x)\nonumber\\
&  =\frac{1}{4\pi c}\widehat{i}\left(  \frac{1}{\sqrt{1-[(dx/dt)/c]^{2}}}%
\frac{dx}{dt}+\frac{1}{\sqrt{1-[(dX/dt)/c]^{2}}}\frac{dX}{dt}\right)
\frac{2\pi K}{c^{2}}\left(  \frac{4\pi K}{c}\right)  Ll(X-x)
\end{align}
\ \ In the quasi-static limit, we drop the terms in $[(dx/dt)/c]^{2}$ and
rewrite the expression for $\mathbf{P}$ in terms of the constant magnetic flux
$\Phi$, giving\
\begin{align}
c^{2}\mathbf{P}  &  \mathbf{=}\frac{1}{8\pi}\widehat{i}\left(  \frac{dx}%
{dt}+\frac{dX}{dt}\right)  \frac{\Phi^{2}l}{L(X-x)}\nonumber\\
&  =\widehat{i}\frac{1}{8\pi}\frac{\Phi^{2}l}{L(X-x)^{2}}\left(
(X-x)\frac{dx}{dt}+(X-x)\frac{dX}{dt}\right)  \text{ \ \ \ \ \ (quasi-static)}%
\end{align}
If the masses $m$ and $M$ of the plates supporting the current sheets are
regarded as ineligible, the position of the center of energy is at the middle
of the solenoid volume%
\begin{equation}
U\overrightarrow{\mathcal{X}}=\widehat{i}\left(  \frac{1}{8\pi}B^{2}%
Ll(X-x)\right)  \frac{(x+X)}{2}=\widehat{i}\frac{1}{16\pi}\frac{\Phi
^{2}l(x+X)}{L(X-x)}%
\end{equation}
Then the time-rate-of-change of the energy times the center of energy gives%
\begin{equation}
\frac{d}{dt}(U\overrightarrow{\mathcal{X}})=\frac{d}{dt}\left(  \widehat
{i}\frac{1}{16\pi}\frac{\Phi^{2}l(x+X)}{L(X-x)}\right)  =\widehat{i}\frac
{1}{8\pi}\frac{\Phi^{2}l}{L(X-x)^{2}}\left(  X\frac{dx}{dt}-x\frac{dX}%
{dt}\right)
\end{equation}
The position-weighted power required on the left-hand side of Eq. (14) is%

\begin{equation}
(\mathbf{F}_{extx}\cdot\mathbf{v})\widehat{i}x+\mathbf{F}_{extxX}%
\cdot\mathbf{V})\widehat{i}X=\widehat{i}\left(  \frac{1}{8\pi}\frac{\Phi^{2}%
l}{L(X-x)^{2}}\frac{dx}{dt}\right)  x+\widehat{i}\left(  \frac{-1}{8\pi}%
\frac{\Phi^{2}l}{L(X-x)^{2}}\frac{dX}{dt}\right)  X
\end{equation}
Now combining Eqs.\ (30) and (32), we see that the sum of the right-hand sides
matches the right-hand side of Eq. (33). \ Indeed the quasi-static expansion
of a solenoid satisfies the relativistic law \ Eq. (14) for the center of energy.

\subsubsection{Two Point Charges at Rest}

The final quasi-static example involves two point charges, one of mass $m$
charge $q$ and the other of mass $M$ and charge $Q$, both at rest in some
inertial frame. \ Again the analysis involves both electromagnetic field
energy and also electromagnetic field momentum as these charges are displaced
quasi-statically from $\mathbf{r}$ to $\mathbf{r}^{\prime}$ and from
$\mathbf{R}$ to $\mathbf{R}^{\prime}$ respectively. \ In the limit of
quasi-static motion, there is no radiation emission on changing the
electrostatic configuration and so the center-of-energy theorem can be
verified exactly.

Here again, the only interesting aspects of the conservation laws involve the
energy and the center-of-energy. \ The external forces needed to move the
charges quasi-statically simply balance the electrostatic forces between the
charges%
\begin{equation}
\mathbf{F}_{extm}=\frac{qQ(\mathbf{R-r)}}{\left\vert \mathbf{R-r}\right\vert
^{3}}=-\mathbf{F}_{extM}%
\end{equation}
while the total energy is the rest-mass energy plus the electrostatic energy
\begin{equation}
U=mc^{2}+Mc^{2}+\frac{qQ}{\left\vert \mathbf{R-r}\right\vert }%
\end{equation}
The energy conservation law (10) in the quasi-static limit takes the familiar
form
\begin{equation}
\mathbf{F}_{extm}\cdot\mathbf{v+F}_{extM}\cdot\mathbf{V=\frac{qQ(\mathbf{R-r)}%
}{\left\vert \mathbf{R-r}\right\vert ^{3}}\cdot(v-V)=}\frac{d}{dt}\left(
\frac{qQ}{\left\vert \mathbf{R-r}\right\vert }\right)  =\frac{dU}{dt}%
\end{equation}
where $\mathbf{v=}d\mathbf{r/}dt$ and $\mathbf{V=}d\mathbf{R}/dt$, and
$U=qQ/\left\vert \mathbf{R-r}\right\vert $ is the electrostatic energy
associated with the two point charges. \ Although this law for energy
conservation is familiar, the relativistic law (14) for the center of energy
is not. \ The center of the electromagnetic energy, by symmetry or by direct
integration of the interference energy between the point-charge fields
$[1/(8\pi)]%
{\displaystyle\int}
d^{3}r\,\mathbf{r\,}2\mathbf{E}_{m}\cdot\mathbf{E}_{M}$, is located half-way
between the two charges so that the energy times the center of energy is given
by%
\begin{equation}
U\overrightarrow{\mathcal{X}}=mc^{2}\mathbf{r+}Mc^{2}\mathbf{R}+\frac
{qQ}{\left\vert \mathbf{R-r}\right\vert }\left(  \frac{\mathbf{r+R}}%
{2}\right)
\end{equation}
The evaluation of the position-weighted power on the left-hand side of (14)
involves%
\begin{equation}
(\mathbf{F}_{extm}\cdot\mathbf{v)r+(F}_{extM}\cdot\mathbf{V)R=}\frac
{qQ}{\left\vert \mathbf{R-r}\right\vert ^{3}}\{[(\mathbf{R-r)\cdot
v]r-[}(\mathbf{R-r)\cdot V]R\}}%
\end{equation}
In the low-velocity limit appropriate for quasi-static changes, the linear
momentum of two point charges is given by the sum of the mechanical linear
momentum and the linear momentum in the electromagnetic field\cite{ColeVan}
\begin{equation}
\mathbf{P}\cong m\mathbf{v}+m\mathbf{V}+\frac{qQ}{2c^{2}\left\vert
\mathbf{R-r}\right\vert }\left(  \mathbf{v+V+}\frac{[(\mathbf{R-r)\cdot
v](\mathbf{R-r)}+[}(\mathbf{R-r)\cdot V]}(\mathbf{R-r)}}{\left\vert
\mathbf{R-r}\right\vert ^{2}}\right)
\end{equation}
For the quasi-static displacement, the time rate of change of the energy times
the center of energy in (37) is%
\begin{equation}
\frac{d}{dt}(U\overrightarrow{\mathcal{X}})=mc^{2}\mathbf{v}+Mc^{2}%
\mathbf{V+}\frac{qQ}{2\left\vert \mathbf{R-r}\right\vert }(\mathbf{v+V)-}%
\frac{qQ}{2\left\vert \mathbf{R-r}\right\vert ^{3}}[(\mathbf{R-r)\cdot
}(\mathbf{V-v)]}(\mathbf{r+R)}%
\end{equation}
Then combining Eqs. (39) and (40), we find%
\begin{equation}
\frac{d}{dt}(U\overrightarrow{\mathcal{X}})-c^{2}\mathbf{P=}\frac
{qQ}{2\left\vert \mathbf{R-r}\right\vert }\left(  \frac{[(\mathbf{R-r)\cdot
v](}2\mathbf{\mathbf{r)}-[}(\mathbf{R-r)\cdot V]}(2\mathbf{R)}}{\left\vert
\mathbf{R-r}\right\vert ^{2}}\right)
\end{equation}
This agrees exactly with the position-weighted power expression on the
right-hand side of Eq. (38). \ Hence indeed the relativistic law for the
center of energy is illustrated in this case; the Coulomb potential between
two point charges fits with the low-velocity limit of electromagnetic theory
so as to give continuous motion for the center of energy under quasi-static
displacements by external forces.

\subsection{Systems Involving Acceleration}

In the examples above, we have tried to illustrate how considerations of
momentum and electromagnetic energy enter into the relativistic law for the
center of energy when treating quasi-static changes of stationary systems.
\ Here we wish to note the role of relativistic energy and momentum for
particles. \ The simplest example seems to be that discussed above in Section
A2 involving a parallel plate capacitor where now the masses $m$ and $M$ of
the plates are no longer treated as negligible and where the external forces
providing a static configuration are removed. \ In this case, the parallel
plates $m$ and $M$ of the solenoid accelerated toward each other under
electrostatic attraction. \ We will verify all of the conservation laws for
the quantities in Eqs. (1)-(4), and we will note just where it is that the
distinction between nonrelativistic and relativistic particle mechanics
becomes important.

The parallel plate capacitor example of Section A2 involves motion along only
the $x$-axis. \ Newton's equations of motion for the plates along the $x$-axis
take the form\cite{Onehalf}%
\begin{equation}
\frac{dp_{m}}{dt}=Q\frac{(E+0)}{2}=\frac{2\pi Q^{2}}{L^{2}}=-\frac{dp_{M}}{dt}%
\end{equation}
where the electrostatic force on each plate is due to the average field across
the plate or is regarded as due to the electric field due to the other plate.
\ In the approximation of large parallel plates with small separation, there
is no magnetic field present even if the plates are moving with finite
velocity, so that there is no electromagnetic linear momentum for the system.
\ Therefore the system linear momentum is simply the mechanical momentum of
the particles%
\begin{equation}
\mathbf{P}=\widehat{i}p_{m}+\widehat{i}p_{M}%
\end{equation}
The angular momentum about the origin vanishes
\begin{equation}
\mathbf{L}=0
\end{equation}
since the $x$-axis passes through the center of each plate. \ The energy of
the system includes the mechanical particle energies $U_{m}$ and $U_{M}$ and
the energy in the electric field $\mathbf{E}=\widehat{i}4\pi Q/L^{2}$ between
the plates\ \
\begin{equation}
U=U_{m}+U_{M}+U_{em}=U_{m}+U_{M}+\frac{1}{8\pi}E^{2}L^{2}(X-x)=U_{m}%
+U_{M}+\frac{2\pi Q^{2}(X-x)}{L^{2}}%
\end{equation}
The energy times the center of energy for the system is
\begin{equation}
U\overrightarrow{\mathcal{X}}=\widehat{i}\left(  U_{m}x+U_{M}X+U_{em}%
\frac{x+X}{2}\right)  =\widehat{i}\left(  U_{m}x+U_{M}X+\frac{2\pi Q^{2}%
(X-x)}{L^{2}}\frac{(x+X)}{2}\right)
\end{equation}

The conservation laws can easily be verified for this system by using the
equations of motion. \ The system linear momentum is constant in time%
\begin{equation}
\frac{d\mathbf{P}}{dt}=\widehat{i}\frac{dp_{m}}{dt}+\widehat{i}\frac{dp_{M}%
}{dt}=0
\end{equation}
as follows from Eq. (42) since the forces on the plates are equal in magnitude
and opposite in direction. \ The system energy is constant in time%
\begin{equation}
\frac{dU}{dt}=\frac{dU_{m}}{dt}+\frac{dU_{M}}{dt}+\frac{dU_{em}}{dt}=\left(
\frac{dp_{m}}{dt}-\frac{2\pi Q^{2}}{L^{2}}\right)  v+\left(  \frac{dp_{M}}%
{dt}+\frac{2\pi Q^{2}}{L^{2}}\right)  V=0
\end{equation}
as follows from the equations of motion in (42) when multiplied by $v=dx/dt$
and by $V=dX/dt$ and then added. \ Here it is crucial to note that for both
nonrelativistic and relativistic particle energy%
\begin{equation}
\frac{dU_{mech}}{dt}=\frac{d\mathbf{p}_{mech}}{dt}\cdot\mathbf{v}\text{
\ \ \ }%
\end{equation}
Thus for the nonrelativistic kinetic energy
\begin{equation}
\frac{d}{dt}U_{mech-nonrel}=\frac{d}{dt}(\frac{1}{2}mv^{2})=(m\mathbf{v}%
)\cdot\frac{d\mathbf{v}}{dt}=\frac{d\mathbf{(mv})}{dt}\cdot\mathbf{v}%
=\frac{d\mathbf{p}_{mech}}{dt}\cdot\mathbf{v}%
\end{equation}
while for the relativistic energy
\begin{align}
\frac{d}{dt}U_{mech-rel}  &  =\frac{d}{dt}\left(  \frac{mc^{2}}{[1-(v/c)^{2}%
]^{1/2}}\right)  =\frac{m\ }{[1-(v/c)^{2}]^{3/2}}\mathbf{v\cdot}%
\frac{d\mathbf{v}}{dt}\\
&  =\frac{d}{dt}\frac{m\mathbf{v}}{[1-(v/c)^{2}]^{1/2}}\cdot\mathbf{v=}%
\frac{d\mathbf{p}_{mech}}{dt}\cdot\mathbf{v}%
\end{align}
The system angular momentum is a constant at $\mathbf{L}=0$ for all time.
\ Thus all of the conservation laws treated so far, linear momentum, energy,
and angular momentum, have not required the specification of nonrelativistic
or relativistic particle mechanics in this electromagnetic system. \ However,
the relativistic law for the center of energy is different; this involves the
generator of proper Lorentz transformations and it requires a fully
relativistic treatment. \ Thus the time-rate-of-change of the system energy
times the center of energy follows from Eqs. (46) and (49) as%
\begin{align}
\frac{d(U\overrightarrow{\mathcal{X}}\mathbf{)}}{dt}  &  =\widehat{i}\left(
\frac{dp_{m}}{dt}vx+\frac{dp_{M}}{dt}VX+U_{m}v+U_{M}V+\frac{2\pi Q^{2}}{L^{2}%
}(XV-xv)\right) \nonumber\\
&  =\widehat{i}\left(  U_{m}v+U_{M}V+\left[  \frac{dp_{m}}{dt}-\frac{2\pi
Q^{2}}{L^{2}}\right]  vx+\left[  \frac{dp_{M}}{dt}+\frac{2\pi Q^{2}}{L^{2}%
}\right]  VX\right) \nonumber\\
&  =\widehat{i}\left(  U_{m}v+U_{M}V\right)
\end{align}
where the terms in square brackets vanish because of the equations of motion
in (42). \ We obtain the correct relativistic law (7) only provided%
\begin{equation}
\frac{d}{dt}(U\mathcal{X)=}U_{m}v+U_{M}V=c^{2}\mathbf{P}%
\end{equation}
However, as we see from Eq. (43) this requires that $U_{m}v=c^{2}p_{m}$ and
$U_{M}V=c^{2}p_{M}$. \ This is not true for nonrelativistic particle energy
and momentum. \ It is true only for the relativistic mechanical energy and
momentum where%
\begin{equation}
U_{mech-rel}=\frac{mc^{2}}{[1-(v/c)^{2}]^{1/2}}\text{ \ \ \ and \ \ \ }%
p_{mech-rel}=\frac{mv}{[1-(v/c)^{2}]^{1/2}}%
\end{equation}
\ Thus provided that we use the exact relativistic expressions for particle
energy and momentum as well as the exact results of electromagnetic theory,
the relativistic center-of-energy law (7) is indeed satisfied for a
parallel-plate capacitor where there are no external forces present and the
plates are free to accelerate. \ We notice that the contributions from both
the relativistic mechanical energy and the electromagnetic energy are
absolutely necessary for the validity of the center-of-energy law.

It might seem that the other examples involving a flattened slip-joint
solenoid and two charged particles can be carried over to the situation
allowing accelerations when no external forces are present. \ However, these
extensions fail because the electromagnetic behavior is not correctly treated
for situations of finite velocity and acceleration. \ Although the
electromagnetic field expressions for a capacitor, in the
large-plate-small-separation approximation, do not change at finite velocity
and acceleration, this is not true for the a flattened slip-joint solenoid or
point charges. \ The expressions used in these quasi-static analyses are valid
only in the low-velocity limit and can be extended to the situation of
accelerating particles only in this low-velocity limit. \ The complications
involved are clearly evident in the case of two charged particles. \ The
Darwin Lagrangian\cite{Jackson-Darwin} correctly describes the interaction of
point charges through order $v^{2}/c^{2}$. \ Even in this order, the particle
equations of motion, can be exceedingly complex,\cite{Page&Adams} and beyond
this order one requires the full Maxwell's equations to describe the
electromagnetic field. \ These situations involve radiation emission and do
not seem to lend themselves to simple examples.

\section{Illustrating the Center of Energy Law in Other Inertial Frames}

Since the energy times the center of energy is the generator of proper Lorentz
transformations, it is natural to wish to see the form taken by the examples
in various inertial frames. \ The example involving the acceleration of the
capacitor plates retains its form under any Lorentz transformation along the
$x$-axis. \ \ Indeed, the electric field between the plates remains
$\mathbf{E=}$ $\widehat{i}4\pi Q/L^{2}$ in any such inertial frame and the
expressions for the mechanical energy and momentum are unchanged so that the
entire analysis is identical for any such Lorentz-transformed frame.
\ However, if a Lorentz transformation is made in another direction, then the
situation becomes distinctly more complicated. \ The parallel plate capacitor
requires forces of constraint for its stability. \ Provided these forces of
constraint do no work in a Lorentz-transformed frame, they will not disrupt
the conservation laws, just as they did not in our calculations above.
\ However, in any inertial frame where the forces of constraint do work, there
must be a flow of energy, and hence also of momentum, which invalidate any
conservation laws which do not take account of these flows.\cite{mass-energy}
\ The parallel plates in our examples have finite extent and therefore must
have forces of constraint in the $y$- and $z$-directions, which forces prevent
the charged plates from flying apart. Thus our conservation analysis will hold
in any inertial frame moving with finite velocity in the $x$-direction since
the forces in the $y$- and $z$-directions do no work. The examples involving
the flattened slip-joint solenoid and two charged particles relatively at rest
also require \ forces of constraint which must be analyzed
carefully.\cite{energy-momentum}

\section{DISCUSSION}

Nonrelativistic mechanics is invariant under the group of Galilean
transformations. \ Electrodynamics is invariant under the Poincare group.
\ However, often nonrelativistic particle mechanics is joined with Maxwell's
electromagnetic theory in describing physical phenomena. \ Indeed, in classes
where the elementary examples above are assigned as homework, students
invariably use nonrelativistic equations of particle motion unless explicitly
required to calculate with the relativistic forms. \ Using nonrelativistic
equations of particle motion, students have no trouble with the conservation
laws for linear momentum, angular momentum, and energy. \ Clearly both
nonrelativistic particle dynamics and electromagnetism contain these
conservation laws, and the examples involve simply the transfer of these
quantities from one system to the other through forces. \ It is only in the
invariance of the velocity of the center of energy that we become aware that
Poincare invariance enforces strong restrictions on the theory.
\ Nonrelativistic particle equations of motion fail to yield the invariant
motion of the center of energy when electromagnetic energy and particle
kinetic energy are included. \ 

The three examples of the relativistic conservation laws which we have given
here all involve classical electromagnetism which is invariant under the
Poincare group. \ The example of the accelerating plates of a parallel-plate
capacitor illustrates that mixtures of nonrelativistic and relativistic
physics still lead to the conservation laws for linear momentum, angular
momentum, and energy while only fully relativistic systems satisfy the law for
the center of energy. \ Calculation of the center-of-energy motion forces us
to notice the distinction between relativistic physics and the alternatives.
\ Within relativistic physics, it is not at all clear that particles\ can
interact through any arbitrary potential function $V(\left\vert \mathbf{r-R}%
\right\vert )$; the $1/r$ Coulomb or Kepler potential appears as part of the
relativistic theories of electromagnetism and gravitation. \ Indeed it seems
fascinating that the generator of the O(4) symmetry associated with the
Runge-Lenz vector of the nonrelativistic $1/r$ Kepler problem\cite{Goldstein}
is precisely the nonrelativistic limit of the generator $U\overrightarrow
{\mathcal{X}}$ for proper Lorentz transformations obtained from the Darwin
Lagrangian for the $v^{2}/c^{2}$-interaction of two charged
particles.\cite{Dahl}

The relativistic conservation laws associated with Poincare invariance require
the use of relativistic physics for both the interactions and the mechanical
energy and momentum.\cite{Rohrlich} \ However, both the text book and research
literature in physics contain many examples where nonrelativistic and
relativistic aspects are mixed together. \ This arrangement maintains the
conservation laws of linear momentum, energy, and angular momentum, but not
the relativistic law for the center of energy. \ For the most part this does
not lead to significant difficulties \ in one-step calculations when the
particle mechanics is taken as nonrelativistic in the presence of fixed
electromagnetic fields and only the particle motion is of
interest.\cite{mix-example} \ However, there are multi-step calculations where
the charged particles respond with nonrelativistic motion to electromagnetic
fields and in turn the electromagnetic fields arising from the
nonrelativisticly-moving particles are of interest; these calculations lead to
questionable conclusions. \ Thus, for example, the Aharonov-Bohm phase shift
involves the $v^{2}/c^{2}$-interaction of a point charge and a solenoid; yet
the response of the solenoid to the charged particle's fields is often treated
using nonrelativistic physics.\cite{Tassie} One instance of the paradoxical
and erroneous descriptions which can arise from such\ a treatment of the
charged particle-solenoid interaction is discussed by Coleman and Van
Vleck;\cite{ColeVan} other discussions have also been given.\cite{BoyerAB} \ A
second example involves the scattering of random classical radiation by a
mechanical scatterer so as to obtain the equilibrium spectrum corresponding to
thermal (blackbody) radiation. \ It is common practice\cite{VVBoyer} to use
nonrelativistic mechanical behavior for the scattering charges despite the
fact that the electromagnetic fields arising from the nonrelativisticly-moving
particles are of crucial interest in obtaining radiation equilibrium. \ In
some instances,\cite{Blanco} relativistic particle mechanics has been combined
with nonrelativistic potential functions in an attempt to discuss classical
radiation equilibrium. \ In all these instances, the relativistic
center-of-energy law is violated because the systems do not satisfy Poincare
invariance. \ Yet relativistic transformations are clearly crucial in
understanding blackbody radiation since the Planck spectrum can be obtained by
Lorentz transformations associated with uniform (proper) acceleration through
Lorentz-invariant zero-point radiation.\cite{Unruh}

In this article we have given several elementary examples of the relativistic
center-of-energy law. \ There are very few examples of the law presented in
the text book literature and the relativistic restrictions associated with the
law seem to be unnoticed in some of the physics research literature. \ Thus a
century after Einstein's striking work on special relativity, there are still
elementary aspects of Lorentz invariance which go unmentioned in the text
books and unappreciated in the research literature.


\begin{thebibliography}{99}                                                                                               %


\bibitem {ColeVan}S. Coleman and J. H. Van Vleck, "Origin of `Hidden Momentum
Forces' on Magnets," Phys. Rev. \textbf{171}, 1370-1375 (1968).

\bibitem {CofE}Perhaps in part because there are so few simple examples
involving the center of energy, there is a variety of terminology in the
literature. \ Here we have chosen to speak of the "center of energy,"
following the usage of Coleman and Van Vleck in Ref. 1. \ However, E. F.
Taylor and J. A. Wheeler in \textit{Spacetime Physics} (Freeman, San
Francisco, 1966), p. 143, speak of the "center of mass" with the understanding
that "mass" means "mass-energy" as befits a relativistic theory. \ L. D.
Landau and E. M. Lifshitz in \textit{The Classical Theory of Fields}%
,\textit{\ }4th ed\textit{.} (Pergamon, New York, 1985), p. 168, speak of the
"center of inertia." \ J. L. Anderson, "\textit{Principles of Relativity
Physics}" (Academic Press, New York 1967), p. 207-208 writes, "We can call
$\mathbf{Z}$ the center of energy of the system of particles, in analogy to
the center of mass as defined in Newtonian mechanics. \ According to
Eq.(7-5.7), this point moves like a free particle with a velocity $\mathbf{V}%
$, given by Eq. (7-5.9) \ If $P^{\mu}P_{\mu}=M^{2}>=0,$ we can always perform
a mapping so that $P^{\mu}=(M,0).$ \ The corresponding reference frame is
called variously, the center of mass, or center of momentum, or center of
energy frame. \ We prefer the latter terminology. \ In this frame
$\mathbf{V}=0."$ \ All of these designations mean the same thing. \ Here the
"center of energy" terminology has been used so as to impress upon the reader
that this is a change in point of view from the nonrelativistic "center of
(rest) mass" concept where there is no role for electromagnetic energy or even
particle kinetic energy and only particle rest masses are involved.

\bibitem {Griffiths}See, for example, D. J. Griffiths, \textit{Introduction to
Electrodynamics}, 3rd ed. (Prentice Hall, Upper Saddle River, NJ. 1999).
\ Chapter 8 is devoted to conservation laws in electromagnetic theory--
including conservation of charge, energy, linear momentum, and angular
momentum. \ There is no mention of the invariant motion of the center of
energy. \ Indeed, it was Griffiths' clear organization of the conservation
laws which made me acutely aware of the absence of the last conservation law
of Poincare invariance.

\bibitem {Examples}I an not aware of any simple examples of the
center-of-energy theorem in electromagnetism textbooks. \ L. D. Landau and E.
M. Lifshitz in \textit{The Classical Theory of Fields}, 2nd ed. (Pergamon,
Oxford, 1962), p. 194, give as a problem the determination of the "center of
inertia" for a collection of interacting point charges. \ This problem is
repeated on page 168 of the 4th edition listed in Ref. 2 above. \ J. D.
Jackson in \textit{Classical Electrodynamics} 2nd ed. (Wiley, New York, 1975),
p. 617, has problem 12.16 to derive the uniform motion of the "center of mass"
for an arbitrary, localized distribution of source-free electromagnetic
fields. \ \ The question is repeated as problem 12.19 in the 3rd edition of 1999.

\bibitem {Einstein}The basic idea appears in the early work of A. Einstein,
"Prinzip von der Erhaltung der Schwerpunktsbewegung und die Tr\"{a}gheit der
Energie," Annalen der Physik \textbf{20}, 626-633 (1906). \ It is also given
by E. Bessel-Hagen, "\"{U}ber die Erhaltungss\"{a}tze der Elektrodynamick,"
Mathematisch Annalen \textbf{84}, 259-276 (1921). \ Bessel-Hagen analyzes the
conservation laws associated with the conformal group satisfied by Maxwell's equations.

\bibitem {Poynting}See any standard text on electromagnetic theory; for
example, Griffiths' Section 8.1.2 "Poynting's Theorem" in Ref. 3. \ 

\bibitem {unaware}I am not aware of any place where the center-of-energy law
with external forces appears in the physics literature.

\bibitem {Onehalf}The forces on each plate can be regarded as due to the
average electric field across the plate, or as due to the electric field of
the other plate, or as due to the pressure of the electromagnetic field.
\ (See, for example, D. J. Griffiths in Ref. 3, p. 102, Eq. (2.50), or E. M.
Purcell, \textit{Electricity and Magnetism}, 2nd ed. (McGraw-Hill, New York,
1985), pp. 30 and 31.)

\bibitem {solenoid}See, for example, T. H. Boyer, "Electric and magnetic
forces and energies for a parallel-plate capacitor and a flattened, slip-joint
solenoid," Am. J. Phys. \textbf{69}, 1277-1279 (2001).

\bibitem {Jackson-Darwin}See, for example, J. D. Jackson, \textit{Classical
Electrodynamics, 2nd ed. (Wiley, New York 1975), }Section 12.7 "Lowest-Order
Relativistic Corrections to the Lagrangian for Interacting Charged Particles,
the Darwin Lagrangian."

\bibitem {Page&Adams}See, for example, the fields given by L. Page and N. I.
Adams, "Action and reaction between moving charges," Am. J. Phys. \textbf{13},
141-147 (1945). \ These electromagnetic fields follow from the Darwin Lagrangian.

\bibitem {mass-energy}See, for example, the discussion in the introduction of
the article by T. H. Boyer, "Example of mass-energy relation: Classical
hydrogen atom accelerated or supported in a gravitational field," Am. J. Phys.
\textbf{66}, 872-876 (1998).

\bibitem {energy-momentum}See, for example, T. H. Boyer,
"Lorentz-transformation properties for energy and momentum in electromagnetic
systems," Am. J. Phys. \textbf{53}, 167-171 (1985).

\bibitem {Goldstein}See, for example, H. Goldstein, \textit{Classical
Mechanics 2nd ed.} (Addison-Wesley , Reading, Massachusetts 1980), Section 3-9
"The Laplace-Runge-Lenz Vector."

\bibitem {Dahl}J. P. Dahl, "Physical origin of the Runge-Lenz vector," J.
Phys. A: Math. Gen. \textbf{30}, 6831-6840 (1997).

\bibitem {Rohrlich}F. Rohrlich in \textit{Classical Charged Particles}
(Addison-Wesley, Reading, MA 1965), p. 210, emphasizes that the combination of
nonrelativistic particle mechanics and electromagnetic fields is
"inconsistent" in the sense that the combination satisfies neither Galilean
invariance nor Lorentz invariance.

\bibitem {mix-example}See, for example, Ref. 3, Example 5.2, where Griffiths
discusses "Cycloid Motion" of a nonrelativistic charged particle in electric
and magnetic fields.

\bibitem {Tassie}See, for example, the calculations by M. Peshkin, I. Talmi,
and L. J. Tassie, "The Quantum Mechanical Effects of Magnetic Fields Confined
to Inaccessible Regions," Ann. Phys. (N.Y.) \textbf{12}, 426-435 (1961),
especially Section V, "A Mechanical Model."

\bibitem {BoyerAB}See, for example, T. H. Boyer, "Classical Electromagnetic
Interaction of a Point Charge and a Magnetic Moment: Considerations Related to
the Aharonov-Bohm Phase Shift," Found. Phys. \textbf{32}, 1-39 (2002).

\bibitem {VVBoyer}See, for example, J. H. Van Vleck, "The Absorption of
Radiation by Multiply Periodic Orbits, and its Relation to the Correspondence
Principle and the Rayleigh-Jeans Law. Part II Calculation of Absorption by
Multiply Periodic Orbits," Phys. Rev. \textbf{24}, 347-365 (1924) and T. H.
Boyer, "Equilibrium of random classical electromagnetic radiation in the
presence of a nonrelativistic nonlinear electric dipole oscillator," Phys.
Rev. \textbf{13}, 2832-2845 (1976).

\bibitem {Blanco}See R. Blanco, L. Pesquera, and E. Santos, "Equilibrium
between radiation and matter for classical relativistic multiperiodic systems.
Derivation of Maxwell-Boltzmann distribution from Rayleigh-Jeans spectrum,"
Phys. Rev. D \textbf{27}, 1254-1287 (1983); "Equilibrium between radiation and
matter for classical relativistic multiperiodic systems II. Study of radiative
equilibrium with Rayleigh-Jeans radiation," Phys. Rev. D \textbf{29},
2240-2254 (1984).

\bibitem {Unruh}P. C. W. Davies, \textquotedblleft Scalar particle production
in Schwarzschild and Rindler metrics," J. Phys. A \textbf{8}, 609-616 (1975);
W. G. Unruh, \textquotedblleft Notes on black-hole evaporation," Phys. Rev. D
\textbf{14}, 871-892 (1976); T. H. Boyer, \textquotedblleft Thermal effects of
acceleration for a classical dipole oscillator in classical electromagnetic
zero-point radiation," Phys. Rev. D \textbf{29}, 1089--1095 (1984); D. C.
Cole, \textquotedblleft Properties of a classical charged harmonic oscillator
accelerated through classical electromagnetic zero-point
radiation,\textquotedblright\ Phys. Rev. D \textbf{31}, 1972--1981 (1985).
\end{thebibliography}
\end{document}